# Impacts of climate change on groundwater droughts by means of standardized indices and regional climate models

Daniele Secci*, Maria Giovanna Tanda, Marco D'Oria, Valeria Todaro, Camilla Fagandini

Department of Engineering and Architecture, University of Parma, Parco Area delle Scienze 181/A, 43124 Parma, Italy

* Corresponding author, daniele.secci@unipr.it

**Abstract**

This paper investigates the impacts of climate change on groundwater droughts making use of regional projections and standardized indices: the Standardized Precipitation Index (SPI), the Standardized Precipitation Evapotranspiration Index (SPEI) and the Standardized Groundwater Index (SGI). The method adopted, using historical precipitation and temperature data and water levels collected in monitoring wells, first investigates the possible correlations between meteorological and groundwater indices at each well. Then, if there is a correlation, a linear regression analysis is used to model the relationships between SGIs and SPIs, and SGIs and SPEIs. The same relationships are used to infer future SGIs from SPI and SPEI projections obtained by means of an ensemble of Regional Climate Models (RCMs), under different climate scenarios (RCP 4.5 and RCP 8.5). This methodology has been applied to data collected in northern Tuscany (Italy) in an area served by a water company, where historical series of daily climate variables (since 1934) and daily records for 16 wells, covering the period 2005-2020, are available. The impacts on groundwater have been computed in the short- (2006-2035), medium- (2036-2065) and long-term (2066-2095). The analysis indicates that, in the historical period and for most of the monitoring wells, there is a good correlation between SGIs and SPIs or SPEIs. The results point out that making use of the SGI-SPI relationships, slight variations in the availability of groundwater are expected in the future. However, in a global warming scenario, the influence of temperature on evapotranspiration phenomena cannot be overlooked and, for this reason, the SGI-SPEI relationships seem more suitable to forecast groundwater droughts. According to these relationships, negative effects on groundwater levels in almost all wells are estimated for the future. For the RCP 4.5 scenario, the largest decline in groundwater level is expected in the medium-term, while for the RCP 8.5 scenario future SGIs will significantly decrease over the long-term. Due to the type of data required and its simplicity, this methodology can be applied to different areas of interest for a quick estimate of groundwater availability under climate change scenarios.

# 1 Introduction

Climate change is one of the most addressed issues of the twenty-first century as its negative impacts on the environment are increasingly evident (e.g. Jiménez Cisneros et al., 2015). Therefore, environmental protection is a key concern for this century and, certainly, aquifers cannot be left behind for their significant contribution to water supply, irrigation and industrial needs.

In the fifth Assessment Report of the Intergovernmental Panel on Climate Change IPCC, 2014), the assessment of future climate is linked to different projections of anthropogenic greenhouse gases (GHG) emissions, which are the key drivers of increasing global warming. In particular, the IPCC bases its findings on four different 21st century pathways of GHG emissions and atmospheric concentrations, air pollutant emissions and land use: the Representative Concentration Pathways (RCPs) or scenarios, namely RCP 2.6, RCP 4.5, RCP 6 and RCP 8.5 (IPCC, 2013; Moss et al., 2010). To simulate the future climate variables, as a function of the four scenarios, Global Climate Models (GCMs) have been developed by several research centers within the World Climate Research Programme in the Coupled Model Inter-comparison Project, Phase 5 framework (CMIP5 – Taylor et al., 2012).

However, the GCM resolution (100÷500 km) may not be accurate enough to infer reliable projections at regional scale; for this reason, dynamic downscaling techniques have been developed to obtain Regional Climate Models (RCMs), which increase the GCM resolution up to 10÷50 km. In Europe, MED-CORDEX (Ruti et al., 2016) and EURO-CORDEX (Jacob et al., 2014) represent two of the most important initiatives for the simulation of regional climate data. Despite that, to be used on medium-small scale basins for climate change impact studies, the raw RCM outputs need a bias correction process (Teutschbein and Seibert, 2012). In addition, to assess the uncertainty of the results, it is suggested to use an ensemble of climate models (i.e. different GCM-RCM combinations, D'Oria et al., 2018b).

Investigating the impacts of climate change on groundwater resources is not an easy task. Typically, a complex numerical model is required that involves the subsoil description, the conceptualization of the aquifer system, boundary conditions, and recharge and withdrawal rates. Even if a calibrated model is available, simulating future conditions is challenging and the computational burden can be remarkably high, forcing users to limit the number of periods and scenarios to be analyzed. To overcome these problems, surrogate models have been proposed (Razavi et al., 2012; Asher et al., 2015; Rajaee et al., 2019) as a computationally efficient alternative to numerical models, mainly with the aim at helping in the management and decision processes concerning groundwater resources.

In recent years, many authors have investigated the possible relationships between the groundwater levels, observed in monitoring wells, and the main climate variables, such as antecedent precipitation and temperature. A common approach to explore these links is to use standardized indices (see e.g. Khan et al., 2008; Bloomfield and Marchant, 2013; Kumar et al., 2016; Leelaruban et al., 2017; Soleimani Motlagh et al., 2017; Van Loon et al., 2017; Uddameri et al., 2019; Guo et al., 2021). The main indices widely adopted to monitor and quantify droughts worldwide are the standardized precipitation (SPI) and precipitation-evapotranspiration (SPEI) indices for the meteorological variables, and the standardized groundwater index (SGI) for the aquifers. SPI (McKee et al., 1993) is obtained by processing cumulative precipitation at different time windows of consecutive months; SPEI (Vicente-Serrano et al., 2010) is computed on the so-called "useful precipitation", i.e. the difference between precipitation and evapotranspiration; and SGI (Bloomfield and Marchant, 2013) concerns the groundwater level in monitoring-wells. International portals, containing the maps of these indices updated in real time (EDO, 2021; ISPRA, 2021; CNR IBE, 2021) are accessible to different users such as government, public and private agencies and irrigation authorities or agricultural associations to help in decision making.

Khan et al. (2008) investigated the degree of correlation between the SPI and the fluctuations in shallow groundwater levels in the Murra-Darling Basin in Australia. The overall results showed that the SPI correlates well with fluctuations in groundwater table, however, the correlation coefficients resulted lower for areas where irrigation practices are remarkable and the groundwater recharge has complex characteristics. The precipitation accumulation periods that present the best correlation with groundwater levels are different in each analyzed subregion. The authors claimed that the correlation between SPI and groundwater levels can be adopted as a method of relating climatic impacts on water tables.

Bloomfield and Marchant (2013) analyzed the correlation between SPIs and SGIs at 14 sites across the UK. In particular, it was shown that the computation of SGI presents new challenges on the definition of a suitable statistical distribution of the monthly groundwater levels, presenting a dependence on local peculiarities. A strong and evident relationship between SPIs and SGIs was identified, even if the authors highlighted that hydrological processes vary in space and depend on multiple driving forces, not only on meteorological conditions.

Kumar et al. (2016) analyzed groundwater levels and precipitation records at several sites in Germany and the Netherlands; the dependence of SGI on SPI was investigated. The authors found that a variable precipitation accumulation period over 3-24 months is needed to temporally align SPI and SGI at both local and regional scale. This reflects the smoothed response of groundwater to precipitation signals. The correlation between the considered indices decreases using a uniform accumulation period for computing SPI over the entire domain; therefore, an a priori selection of the SPI accumulation period leads to inadequate characterization of groundwater droughts. Overall for the analyzed areas, the authors claimed the limited applicability of the SPI as a proxy for groundwater droughts; SPEI that accounts for temperature is better suited for drought studies under global warming conditions.

Leelaruban et al. (2017) analyzed groundwater level data from wells located in Central US. In particular, the monthly median depth of the water level from the land surface has been correlated with different meteorological indices, including SPI with accumulation periods varying from 6 to 24 months. The authors found that SPI24 correlates best with the groundwater levels and showed how this index can be used for a quick assessment of groundwater droughts. The relationships between drought and aquifer levels are region-specific and therefore ad-hoc studies are required.

Soleimani Motlagh et al. (2017) investigated groundwater drought in the Aleshtar Plain (Iran) using hierarchy and K-means clustering. They calculated the correlation between SPI and SGI for different clusters, finding that the maximum correlation is achieved using different precipitation accumulation periods for each cluster. The magnitude of the correlation coefficient can be variable among the clusters.

Van Loon et al. (2017) reconstructed the groundwater drought occurred in central and eastern Europe in 2015, analyzing the relationship between SGI and SPEI in a reference period (1958-2013). At first, the link between SGI and SPEI was used to establish the spatially varying optimal accumulation period, highlighting a wide accumulation range (1 to 48 months) over the region.

Then, the SGI-SPEI relationships were used to calculate the SGIs for the year 2015. The authors underlined the importance of using a spatially variable accumulation period over large areas.

Uddameri et al. (2019) discussed the possible use of SPI as a surrogate index of the groundwater drought. They analyzed the link between SPI and SGI for the Edwards Aquifer, Texas. Although the two indices were statistically correlated, the frequency at which both were concurrently in the drought state was lower than 50%. According to the authors, this indicates that SPI could be used only for a qualitative prediction of the groundwater drought. However, using SPI to impose drought restrictions is consistent with the precautionary principle.

Guo et al. (2021) investigated the groundwater droughts using the SGI obtained from the data of four monitoring wells located in Georgia, Massachusetts, Oklahoma and Washington. The authors highlighted that the groundwater droughts vary for different areas due to agricultural and human activities; moreover, duration and severity of droughts in the same area also vary at different time scales. The cross-correlation between SGI and SPI was analyzed to find the time delay between meteorological and groundwater droughts.

Climate models give the opportunity to evaluate SPIs and SPEIs also for future scenarios and to detect the occurrence of drought events, their frequency, intensity and duration (Stagge et al., 2015a), comparing them with the historical data. Stagge et al. (2015a) analyzed historical and future SPIs computed from observed precipitation and RCM data. The results obtained for the future period show significant increases in frequency and severity of meteorological droughts in the Mediterranean region, thereby exacerbating their impacts. On the contrary, the evaluations for northern Europe point out a less frequency and severity of droughts since an increase in precipitation is generally detected. Osuch et al. (2016) investigated possible future climate change effects on dryness conditions in Poland using SPIs based on RCM data. Great attention was given to the bias correction of the RCMs, in order to obtain a good reproduction of the historical precipitation.

Furthermore, using the modified Mann-Kendall test, an analysis of the SPI trends was performed employing the Sen's method to calculate the trend slope. In general, this study confirmed the results of Stagge et al. (2015a), highlighting a difference between the climatic projections obtained from the various RCMs.

In this study, we address the use of historical relationships between meteorological and groundwater indices in combination with regional climate model data to infer the impacts of climate change on groundwater. The method adopted, on the basis of the available historical data (precipitation, temperature and groundwater levels), first investigates the correlation between SGIs and SPIs and SGIs and SPEIs at each monitoring well, using different accumulation periods for the climate variables. Then, for those monitoring wells with a satisfactory correlation, a linear regression analysis is used to model the relationships between meteorological and groundwater indices. Assuming that the hydrological processes will not change over time, the same regression relationships are applied to future SPI and SPEI projections to infer the impact of climate change on groundwater levels. Future SPIs and SPEIs are obtained by means of an ensemble of RCMs, under different climate scenarios (RCP 4.5 and RCP 8.5).

The novelty of this study lies in the coupling of drought indices and future projections of climate data to obtain a quick estimate of groundwater availability. In fact, even if many studies focus on the relationships between meteorological and groundwater indices, their use in future analysis is still very little investigated. Employing two different meteorological indices (SPI and SPEI) in combination with SGI, allows to highlight the differences in considering only precipitation rather than precipitation-temperature data to analyze the impact of climate change on groundwater resources. In fact, the use of other climate variables other than precipitation in characterizing droughts is an important aspect emphasized by many others (e.g. Vincente-Serrano et al., 2010; Teuling et al., 2013; Kumar et al., 2016).

The procedure has been applied to a regional area located in northern Italy served by a water company where historical daily data of precipitation, temperature and groundwater levels in wells are available.

This paper is organized as follows: in Section 2, the study area and the available data are presented, then the methodologies adopted to compute SPIs, SPEIs and SGIs and the processing of the climate projections are reported. Section 3 shows the main results, which are discussed in Section 4. Conclusions are drawn in Section 5.

## 2 Materials and methods

### 2.1 Study area and available data

The study area, shown in Fig. 1, is located in the northern part of Tuscany (Italy) and covers about 3000 km$^2$. It is the territory served by an Italian water company, interested in evaluating the effect of climate change on water resources. The anthropic occupation of this area has undergone radical changes. Although agriculture has been the main activity in the last century, it is presently in decline and tourism represents the main source of income (Pranzini et al., 2019). In the last twenty years, the percentage of land used for agricultural is around 14-16% of the total area, resulting in a quite modest water demand. Natural forests occupy between 55 and 70% of the total area (PTA, 2005).

The area has been already investigated in previous studies (D'Oria et al., 2017; D'Oria et al., 2019) and, in agreement, it has been split according to the water divides of four watersheds (Fig. 1): Magra, Serchio, Coastal Basins and Arno Portion (a portion of the Arno River basin). It was necessary to distinguish the area in basins since they have different characteristics. Table 1 summarizes the annual precipitation and annual mean temperature over the four basins as evaluated in the period 1934-2020.

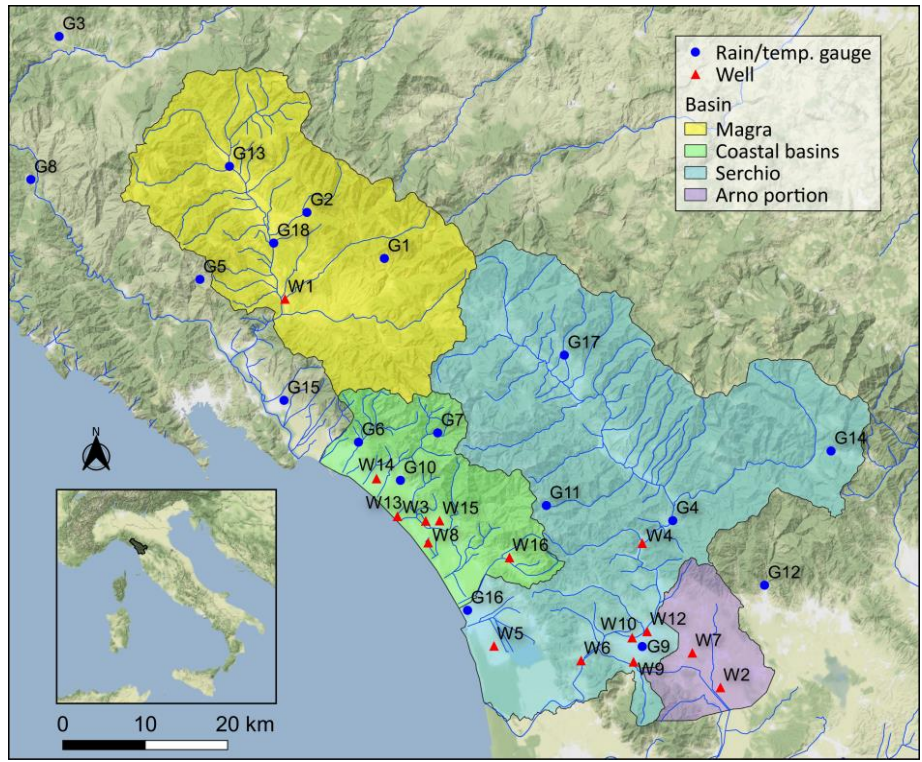

*Fig. 1 - Location of the study area with indication of the climate stations, monitoring wells and river basins.*

*Table 1 - Annual mean temperature and annual precipitation over the basins: average, maximum and minimum values in the period 1934-2020.*

| Annual mean temperature (°C) | MAGRA | COASTAL BASINS | SERCHIO | ARNO PORTION |
|---|---|---|---|---|
| Average | 13.2 | 13.2 | 12.9 | 14.8 |
| Max | 14.8 | 14.8 | 14.3 | 16.0 |
| Min | 11.3 | 11.8 | 11.4 | 13.3 |
| **Annual precipitation (mm)** | **MAGRA** | **COASTAL BASINS** | **SERCHIO** | **ARNO PORTION** |
| Average | 1539 | 1578 | 1536 | 1205 |
| Max | 2608 | 2579 | 2650 | 2039 |
| Min | 810 | 803 | 825 | 444 |

The basin of the Magra River (938 km²) is divided in three different areas: coastal, hilly and mountain portion; the coastal part of the basin is not included in the study area. High spatial variability of the temperature, due to the coastal climate influence, characterizes the hilly area. The inner mountain area presents average winter temperatures close to zero and moderate snow accumulations; high precipitation occurs in the internal areas.

The Coastal basins (383 km²) are located in the area between the Apuan Alps and the Tyrrhenian Sea. The basins are characterized by high precipitation values due to the proximity of the Apuan Alps (maximum altitude 1946 m a.s.l.) to the sea. The most intense rains occur during late spring and late summer, the most persistent one in the autumn; only sporadic and short duration snow occurs due to the high temperature in the winter season.

The Serchio River (1545 km²) has its source in the Apennine area (north of the Province of Lucca) and flows into the Tyrrhenian Sea. The particular position of the basin, longitudinally oriented with the sea, makes the area one of the wettest in Italy, with annual total precipitation exceeding 2500 mm per year on the Apuan hills.

Until the 16th century, the Arno portion area (186 km²) was occupied by swamps and by a lake with an irregular regime draining to Serchio River or Arno River according to the seasonal variations. Then, the zone was reclaimed by means of an artificial channel and the water was addressed to the Arno River. Precipitation is distributed over the year in two periods: between the months of January and May, precipitation is abundant and regular; from October to December, precipitation can be significant and intense but irregularly distributed over time.

In this work, precipitation and temperature data recorded among 18 gauging stations and the piezometric level measurements collected in 16 wells are considered; the climate data extend to the neighboring regions (Liguria and Emilia Romagna regions). The data are published by the Environmental Agency of the three regions (SIR, 2021; ARPAE, 2021; OMIRL, 2021).

The historical daily precipitation and temperature database (years 1934-2012) used in D'Oria et al. (2017) was integrated until 2020. Eighteen precipitation gauges and 14 temperature stations, whose location is plotted in Fig. 1, have been selected to represent the historical climate due to their long period of records; Table 2 shows the type of data recorded and the elevation of each station.

*Table 2 – Type of data and elevation of the precipitation and temperature gauges.*

| ID | Name | Rain gauge | Temp. gauge | Elevation [m a.s.l.] |
|---|---|---|---|---|
| G1 | Arlia | x | x | 460 |
| G2 | Bagnone | x | x | 195 |
| G3 | Bedonia | x | x | 500 |
| G4 | Borgo a Mozzano | x |  | 100 |
| G5 | Calice al Cornoviglio | x | x | 402 |
| G6 | Carrara | x | x | 55 |
| G7 | Casania | x |  | 845 |
| G8 | Cembrano | x | x | 410 |
| G9 | Lucca | x | x | 16 |
| G10 | Massa | x | x | 150 |
| G11 | Palagnana | x | | 861 |
| G12 | Pescia | x | x | 78 |
| G13 | Pontremoli | x | x | 340 |
| G14 | S. Marcello Pistoiese | x | x | 618 |
| G15 | Sarzana | x | x | 26 |
| G16 | Viareggio | x | x | 0 |
| G17 | Villacollemandina | x | | 502 |
| G18 | Villafranca Lunigiana | x | x | 156 |

Daily data from 16 wells are used in this study (Fig. 1 and Table 3); the available data are groundwater levels in m a.s.l. and cover the period 2005-2020. Almost all wells present consistent data time series, except for the S. Pietro a Vico well, which is characterized by few records of the piezometric levels (Table 3) and it was not used for the following analysis.

All the wells considered have been recognized as belonging to underground water bodies that have been classified in terms of the European Directive 2008/105/CE (EU Directive, 2008) and its following national laws D. Lgs. 152/06 (GU, 2006) and D. Lgs. 30/09 (GU, 2009). In the Magra basin, only one monitoring well is available (Bandita 7); it is located in the city of Aulla in the bed aquifer of the Magra River. The Magra groundwater body (21MA010) reaches a depth of a few tens of meters resting on the impermeable sediments of the Rusciniano-Villafranchiano substratum. This aquifer presents a certain lateral continuity along the course of the Magra River and of the main tributaries, with variable thicknesses from the centre to the edges of the plain (D.R. 100, 2010; Regione Toscana, 2021).

Seven monitoring wells are available in the Coastal basin (Table 3); they are dug in the Versilia and Apuan Riviera groundwater body (33TN010; D.R. 100, 2010; Regione Toscana, 2021). It is a multilayer system that presents silt or clayed-silt lenses with good continuity only to a limited extent. Then a direct contact among the aquifer horizons exists on the main part of this water body. The main supply to the groundwater flow comes from the upstream basins and, in particular, from the alluvial fans of the streams (Pranzini et al., 2019).

In the Serchio basin there are six monitoring wells (Table 3). The Decimo well is located in the upper-medium valley of the Serchio River groundwater body (12SE020; Regione Toscana, 2021), which has a depth of 20-30 meters resting on the impermeable sediments of the Pliocene substratum. This phreatic aquifer presents a certain lateral continuity along the course of the Serchio River and of the main tributaries, with variable thickness from the center to the edges of the plain (Regione Toscana, 2021). The other wells are located in the Lucca plain groundwater body – phreatic and Serchio zone (12SE011; Regione Toscana, 2021). The hydrogeological conditions are of a phreatic aquifer.

Two wells are located in the Arno portion basin; they belong to the Lucca plain – Bientina area groundwater body (11AR028; Regione Toscana, 2021). The aquifer is mainly phreatic; only in the southern area a shallow confining layer can be recognized.

*Table 3 – ID, name, reference groundwater body, percentage of available data and ground elevation of the monitoring wells.*

| ID  | Name            | Groundwater body | % data | Elevation m a.s.l. |
|-----|-----------------|------------------|--------|--------------------|
| W1  | Bandita 7       | 21MA010          | 73.4   | 68.00              |
| W2  | Corte Spagni    | 11AR028          | 83.8   | 9.07               |
| W3  | Cugnia          | 33TN010          | 91.7   | 4.00               |
| W4  | Diecimo         | 12SE020          | 60.9   | 65.00              |
| W5  | Flor Export     | 12SE011          | 64.6   | 1.67               |
| W6  | Nozzano         | 12SE011          | 78.6   | 16.43              |
| W7  | Paganico        | 11AR028          | 72.4   | 13.00              |
| W8  | Percorso vita   | 33TN010          | 78.1   | 1.56               |
| W9  | Ronco           | 12SE020          | 79.7   | 11.67              |
| W10 | Salicchi        | 12SE011          | 83.3   | 27.12              |
| W11 | S.Alessio       | 12SE011          | 71.9   | 18.87              |
| W12 | S.Pietro a Vico | 12SE011          | 12.0   | 30.69              |
| W13 | Sat 1           | 33TN010          | 75.5   | 1.50               |
| W14 | Unim            | 33TN010          | 91.7   | 19.91              |
| W15 | Via Barsanti    | 33TN010          | 91.7   | 20.00              |
| W16 | Via Romboni     | 33TN010          | 88.0   | 37.92              |

### 2.1.1 Data compilation: gap filling and interpolation procedures

During data collection, gaps were present within the time series. To fill these blanks and to have a continuous set of observations we used the FAO method (Allen et al., 1998). According to this method, the gaps are filled according to a linear relationship between the data at the considered location and a twin location in which the missing data are available; the data available in the two locations must have a satisfactory correlation. This method was used to fill gaps in groundwater level, precipitation and temperature datasets; as suggested in Allen et al. (1998) a threshold value of 0.7 has been adopted for the correlation coefficient to select twin stations. It is noteworthy that the Bandita7 well, after the FAO filling process, still presents missing data due to the unsatisfactory correlation with the other wells.

Among the 18 climate gauging stations, four have no temperature data. Since this work needs precipitation and potential evapotranspiration data, it is necessary to have contemporary records of temperature and precipitation at the same location. Therefore, once the gaps in the time series were filled, the temperature data were interpolated to the precipitation station locations. For this purpose, in agreement with Moisello (1998), we considered that there is a temperature reduction with increasing ground elevation. Hence, in the recorded period and on a monthly scale, the coefficients $q$ and $m$ of the following linear equation (1) have been determined by means of the ordinary least square (OLS) method applied to the $N$ locations with known temperature $T_j^o$ and elevation $E_j$:

$$T_j^o = q - m \cdot E_j \quad (j = 1, \dots, N). \tag{1}$$

Once estimated the coefficients $q$ and $m$, if (1) is applied to the sites where the temperature record exists, deviations (residuals) can be recognized due to local peculiarities not described by the linear regression. Then, in the estimation of the final temperature $T_i$ in any point of elevation $E_i$, the residuals, weighted with an inverse square distance method, were added to the result of equation (1) giving the following relationship:

$$T_i = q - m \cdot E_i + \sum_{j=1}^{N} \lambda_{i,j} \cdot \varepsilon_j \tag{2}$$

where $\lambda_{i,j}$ is the weight of the $\varepsilon_j$ residual of the temperature values in the $j$ location with known temperature.

### 2.1.2 Future climate projections

Estimates of the future climate in terms of daily precipitation and daily mean temperature have been acquired from an ensemble of 13 RCM models, which are part of the EURO-CORDEX initiative (Jacob et al., 2014). The combinations of GCMs and RCMs adopted in this study are reported in Table 4. The RCM data consist of a historical control period (1950/1970-2005) and a projection period of the climate variables from 2006 until 2100, under different Representative Concentration Pathways (RCPs). In this work, the RCP 4.5 and RCP 8.5 scenarios have been considered. The climate model data have been downscaled at the 18 climate stations and bias corrected with reference to the historical period 1976-2005. In particular, the climate model data (daily precipitation and temperature) have been corrected with the Distribution Mapping method (Teutschbein and Seibert, 2012) so that their cumulative distribution functions, at monthly scale, agree with the ones of the observed data in the chosen historical period. The same correction estimated for the historical period is then applied for the future. For more information and details on the climate models data, the downscaling and the bias correction method for the study area see D'Oria et al. (2017).

*Table 4 – Combination of GCMs and RCMs from the EURO-CORDEX project used in this study.*

|     |           | GCM      |          |           |            |              |
| --- | --------- | -------- | -------- | --------- | ---------- | ------------ |
|     |           | CNRM-CM5 | EC-EARTH | HadGEM2-ES | MPI-ESM-LR | IPSL-CM5A-MR |
| RCM | CCLM4-8-17 | x        | x        | x         | x          |              |
|     | HIRHAM5   |          | x        |           |            |              |
|     | WRF331F   |          |          |           |            | x            |
|     | RACMO22E  |          | x        | x         |            |              |
|     | RCA4      | x        | x        | x         | x          | x            |

## 2.2 Calculation of drought indices

In this section, we first describe the methodologies used to compute the meteorological indices, SPI and SPEI, and the groundwater index, SGI, in the historical period. Then, the methodology to evaluate the relationships between meteorological and groundwater indices in the historical period is presented. Finally, we show how to evaluate the future SGIs on the basis of the SPI and SPEI projections and the previously estimated relationship.

### 2.2.1 Standardized Precipitation Index (SPI)

The Standardized Precipitation Index (SPI) was developed by McKee et al. (1993) and represents a statistical index useful in detecting the severity of meteorological droughts. The computation of SPI requires a long series of monthly precipitation (30 years or more is suggested by the World Meteorological Organization (1987)), accumulated over different time windows of interest (e.g. 1, 3, 6, 9, 12, 24 months). The precipitation values related to a certain month and time window are first fitted to an appropriate probability distribution, which is then transformed into a standard normal distribution. SPI values close to zero indicate precipitation close to the average, positive or negative values indicate abundant or scarce rains; negative values less than -1 denote the occurrence of a meteorological drought.

In the present study, the SPI has been evaluated at station scale on the basis of the long-term precipitation records of the years 1934-1993, assumed as reference period. The probability distribution function (PDF) that usually fits the cumulative precipitation data is the gamma distribution (McKee et al., 1993, Soľáková et al., 2014, Stagge et al., 2015b) and this has been used in this work.

Care must be taken to the so-called "zero precipitation problem". During a season with low precipitation, the accumulated precipitation over short periods (1 or 3 months, generally) can be zero, but the gamma distribution can only handle positive values. Hence, according to Stagge et al. (2015b), the cumulative gamma distribution function was transformed in a piecewise probability distribution as follows:

$$p(x) = \begin{cases} p_0 + (1-p_0)G(x_{p>0}, \boldsymbol{\gamma}) & \text{for } x > 0 \\ p_0 = \dfrac{n_{p=0}+1}{2(n+1)} & \text{for } x = 0 \end{cases} \tag{3}$$

where $p$ is the probability distribution, $p_0$ is the zero precipitation probability, $n_{p=0}$ is the number of zeros occurring in the total data set of $n$ values, $G(x_{p>0}, \boldsymbol{\gamma})$ denotes the Gamma distribution with parameters $\boldsymbol{\gamma}$ and $x$ is one element in the series.

In this study, the distribution function fitted over the reference period was used to calculate the SPIs in more recent years (2005-2020), when the groundwater level data were available. The choice of not extending the reference period until 2020 is related to the fact that, in the study area, the effects of climate change have been detected since the '90s (D'Oria et al., 2017).

Once the SPIs have been computed at each gauging station, we processed them in order to obtain an average value according to the Thiessen polygon method. In particular, we evaluated the average SPIs for each basin and for the total area.

### 2.2.2 Standardized Precipitation-Evapotranspiration index (SPEI)

In hydrological processes, temperature can play a non-negligible role; for this reason, in addition to the SPI, the Standardized Precipitation-Evapotranspiration Index (SPEI) has been considered in this work. The procedure for calculating the SPEI (Vicente-Serrano et al., 2010) is quite similar to that used for the SPI; in this case the reference meteorological variable is the difference between the precipitation and the potential evapotranspiration (PET). In this work, the PET has been evaluated in agreement with the Thornthwaite method (Thornthwaite, 1948) since only mean temperature data were available for the study area.

The gamma distribution used for the SPI no longer accommodates the useful precipitation data, because negative values may occur due to the contribution of the evapotranspiration. According to Stagge et al. (2015b), we used the log-logistic distribution. Once the distribution is fitted, the data are transformed using a standard normal distribution to obtain the SPEI values. The reference period adopted to fit the log-logistic distribution is 1934-1993. The computation of the SPEIs outside the historical reference period may require significant extrapolation of the fitted distribution leading to unreasonable values (Stagge et al. 2015b). In these cases, the SPEIs were limited to the range of the extreme values allowed by the historical distribution.

Again, using the Thiessen polygons and the SPEIs at station scale, their areal averages have been computed for the analyzed basins and the total area.

### 2.2.3 Standardized Groundwater Index (SGI)

As previously mentioned, SGI represents a statistical indicator of the groundwater drought severity, conceptually identical to SPI and SPEI. However, there are significant differences: there is no meaning in the accumulation over a specified period and the distribution of the observed monthly groundwater levels does not conform to the already analyzed PDFs. Some authors used different distributions to analyze the groundwater data such as the plotting position method (Osti et al., 2008) and the kernel non-parametric distribution (Vidal et al., 2010; Bloomfield and Marchant, 2013; Soleimani Motlagh et al., 2017). The plotting position method is very sensitive to the sample size, especially when the number of data is small; for this reason, the kernel non-parametric method is preferred and used in this study. The PDF of the model is the following (Horová et al., 2012):

$$p(x) = \frac{1}{Nh}\sum_{k=1}^{N} K(\frac{x - x_k}{h}) \qquad (4)$$

where $p(x)$ is the probability density function of the variable $x$, $h > 0$ is the bandwidth, $K(x)$ is the kernel function which may be defined in different forms (normal, box, triangle, Epanechnikov) and $x_k$ is a random sample from an unknown distribution. In this study, a Gaussian Kernel is used. Once established the distribution, the normalization procedure for obtaining SGI is identical to the process described for the meteorological indices.

### 2.2.4 Future SPIs and SPEIs

Making use of the outputs of the climate models, SPIs and SPEIs can be evaluated for the historical period and selected future periods. For this aim, the indices have been computed according to the probability distributions used to fit the historical observations in the period 1934-1993.

As highlighted by Stagge et al. (2015a) and Osuch et al. (2016), it can be argued whether the results provided by climate models, once downscaled and bias corrected, well describe the observed standardized indices, like SPIs and SPEIs, in a common historical period. A positive answer gives a certain assurance that the climate models provide reliable predictions of the meteorological indices in the future. In order to investigate this issue, we performed a check on the congruence of the probability distributions of the observed SPIs and SPEIs with the ones obtained from the climate models in the historical period 1976-2005. To this end, we applied the two sample Kolmogorov-Smirnov test, which verifies whether two samples are drawn from the same distribution. The test was applied to the SPIs and SPEIs evaluated, for each station and a certain time window, using the observed and the RCM data. The climate models have been individually tested.

### 2.2.5 Future SGIs

To obtain future projections of SGIs for the study area, first the relationships between SPIs or SPEIs (at different time windows) and SGIs in a historical period must be investigated. To this end, a preliminary correlation analysis was made, based on the Pearson coefficient, on the indices calculated in the period 2005-2020. Investigations were also conducted to verify if potential delays between meteorological and groundwater indices (i.e. shifting backward the SPI or SPEI) may increase their correlation. A threshold for the correlation coefficient equal to 0.6 was adopted to identify an acceptable link between the two indices (Evans, 1996).

For those wells with acceptable correlation, we made use of a regression analysis to establish a simple linear relationship between meteorological indices (SPIs or SPEIs) and SGIs. Then, assuming that the regression equations determined for the historical period hold for the future, they were applied to determine the SGIs according to the future meteorological indices (SPIs or SPEIs). The future analysis were conducted at short- (2006-2035), medium- (2036-2065) and long-term (2066-2095).

# 3 Results

In this section, the main results are summarized. After reporting the SPIs, SPEIs and SGIs computed in the historical period (2005-2020), the correlations between meteorological and groundwater indices are analyzed and their relationships identified. Finally, the future projections of the SGIs are presented.

## 3.1 Historical SPIs, SPEIs and SGIs

Even if SPIs and SPEIs were calculated at station scale, for the sake of brevity, **Fig. 2** shows only those averaged over the basins of interest and for the period of availability of the groundwater levels (2005-2020). The time windows of 6, 9, and 12 months are selected since the highest correlations between meteorological and groundwater indices are in these aggregation periods.

About the variability of the SPIs among the basins, it seems not significant; with reference to the 12-month time window (**Fig. 2a**), all the basins show a drought period that starts in 2005 and ends in 2009. Another remarkable drought is detected from 2012 to 2013; this one is less severe in the Magra basin. Again, limiting the analysis to the 12-month time window, the smallest values are obtained for the Arno portion basin, in 2008; the largest in the Serchio basin in 2012-2013.

The SPEI values (**Fig. 2b**) indicate drought periods similar to those identified by the SPIs; on average, limiting the analysis to the 12-month time window, they result lower in the negative values and moderately higher in the positive ones. The smallest value is obtained for the Arno portion basin in 2012; the largest value for the Serchio basin in 2014.

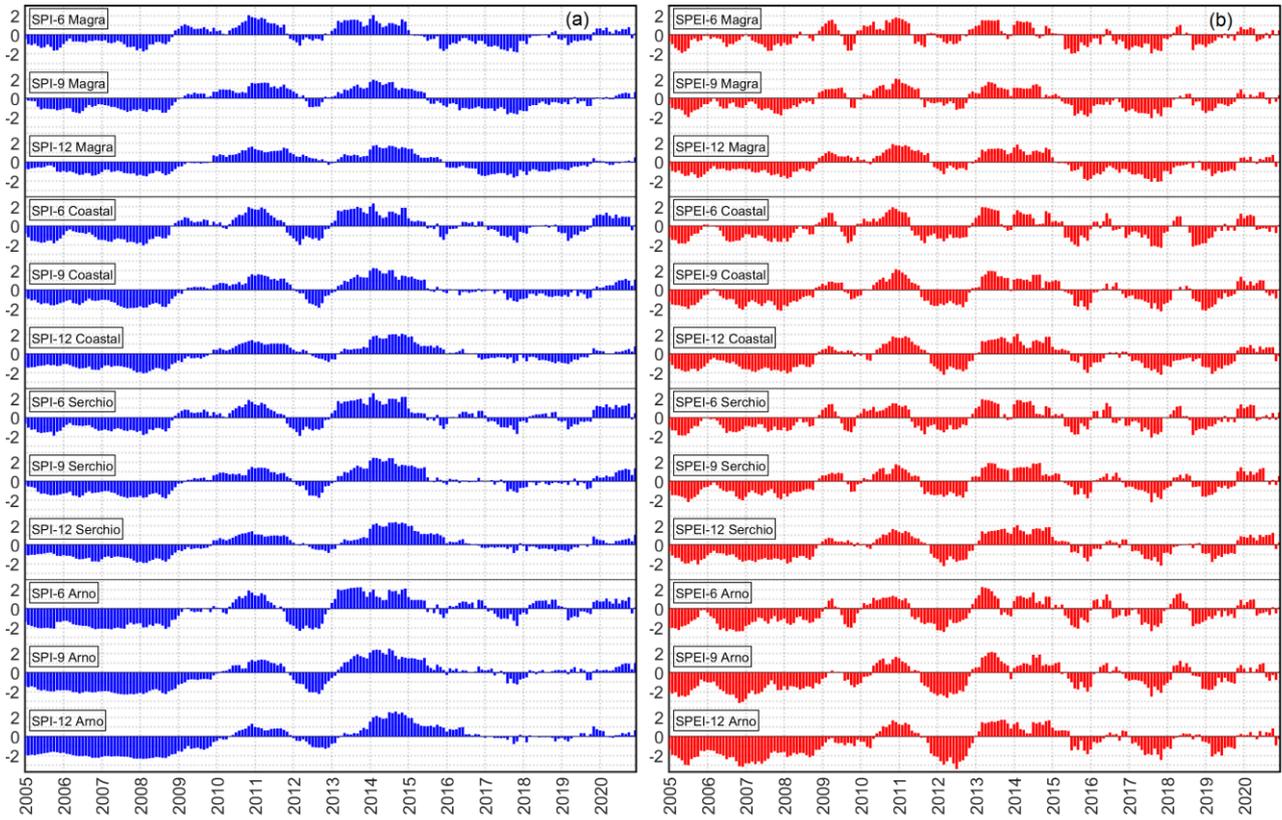

*Fig. 2 – SPIs (a) and SPEIs (b) for the four analyzed basins and time windows of 6, 9 and 12 months.*

The SGIs were calculated for the data collected in the monitoring wells in the period 2005-2020. The SGIs, shown in Fig. 3, detect drought periods similar to those of SPIs and SPEIs for almost all wells. For some wells, in particular Bandita7 (Magra basin), Unim (Coastal basin) and Corte Spagni (Arno basin), some positive or slightly negative values are detected during the drought period of the years 2005-2009. This condition could be due to the influence of external forcing on groundwater. For example, the proximity of the Magra River to the Bandita7 well may influence the groundwater levels, while Unim and Corte Spagni are affected by withdrawals from nearby well fields.

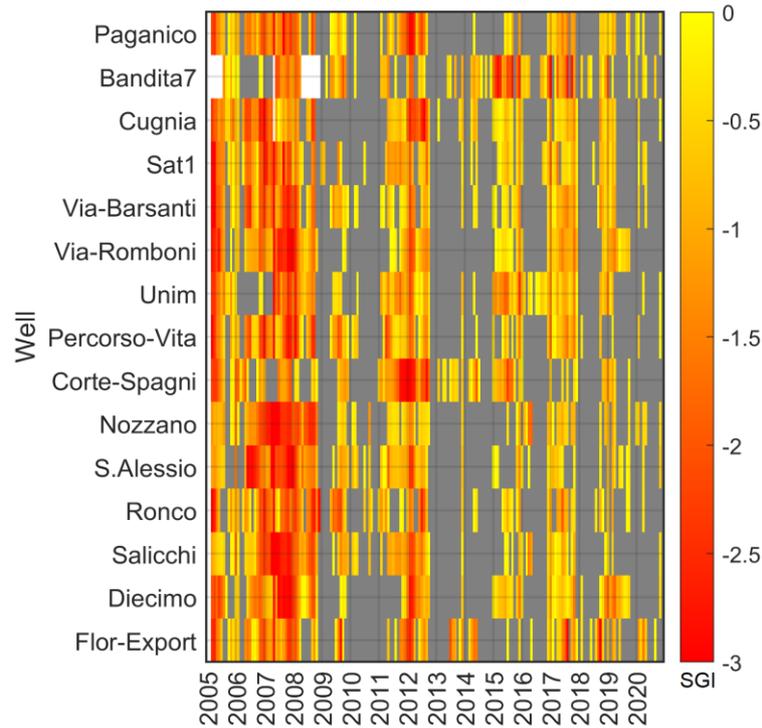

*Fig. 3 - SGI values for the 15 monitoring wells used in this study. The white color indicates missing data, the grey color indicates positive values, while the color scale classifies the negative SGIs.*

### 3.2 Relationships between meteorological and groundwater indices

To recognize potential relationships between meteorological and groundwater indices, we started investigating the correlation between SPIs and SGIs. For each monitoring well, we computed the Pearson correlation coefficient between the SPIs weighted on the corresponding basin and the SGIs. The correlations obtained using the basin weighted SPIs are generally higher than those evaluated with the SPIs weighted over the entire study area; this makes the results more reliable. This is consistent with other literature studies, which highlighted that both the climate and basin characteristics influence the propagation of the precipitation signal to groundwater (e.g. Kumar et al., 2016).

For the correlation analysis, eight time windows (1, 3, 6, 9, 12, 18, 24 and 36 months) were considered and the results are shown in Fig. 4. With reference to the correlation coefficients higher than the chosen threshold (0.6), the SPIs with time windows of six, nine and twelve months are generally better correlated with the SGIs. This behavior was expected considering that the wells are located in shallow aquifers with moderate distance from the ground surface (Kumar et al., 2016). However, some wells present low correlation values for all the considered time windows; this is particularly evident for the Bandita7, Unim and Corte Spagni wells in agreement with the results reported in Section 3.1.

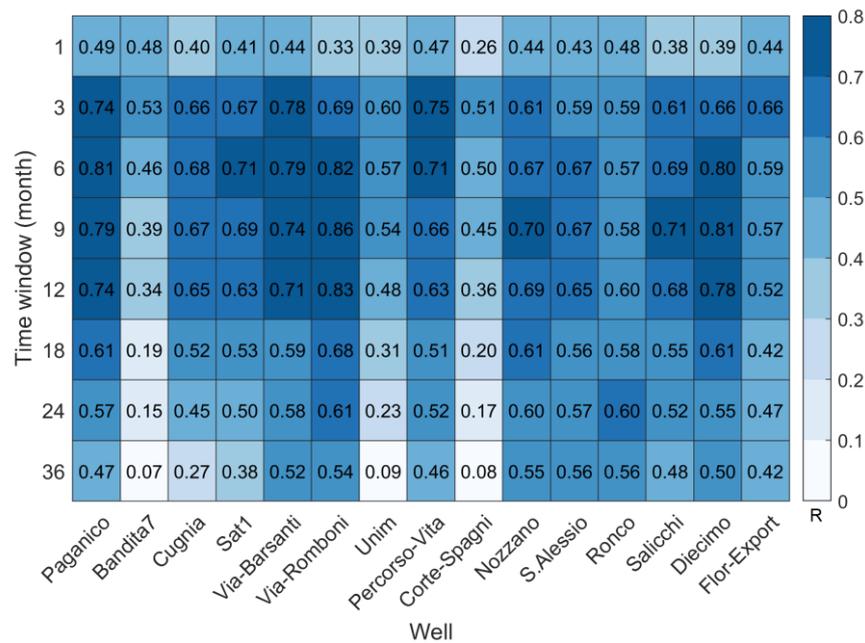

*Fig. 4 - SGI-SPI Pearson correlation coefficients.*

For the following analysis, we will consider only the wells with a correlation coefficient higher than the selected threshold (0.6) and for the 6-, 9-, and 12-month time windows. Ten wells satisfy this condition; they are located in the Arno portion (1 well), Coastal (5 wells) and Serchio (4 wells) basins. As showed by Bloomfield and Marchant (2013), it can be interesting to investigate if a delay (lag) between meteorological and groundwater indices may modify the correlation coefficients, allowing a better alignment between the precipitation and the groundwater signals. The heat maps in **Fig. 5** summarize the computations and show that the highest correlation coefficient is observed for zero-lag. This indicates that, for the study area, the meteorological droughts are aligned to those of the groundwater system.

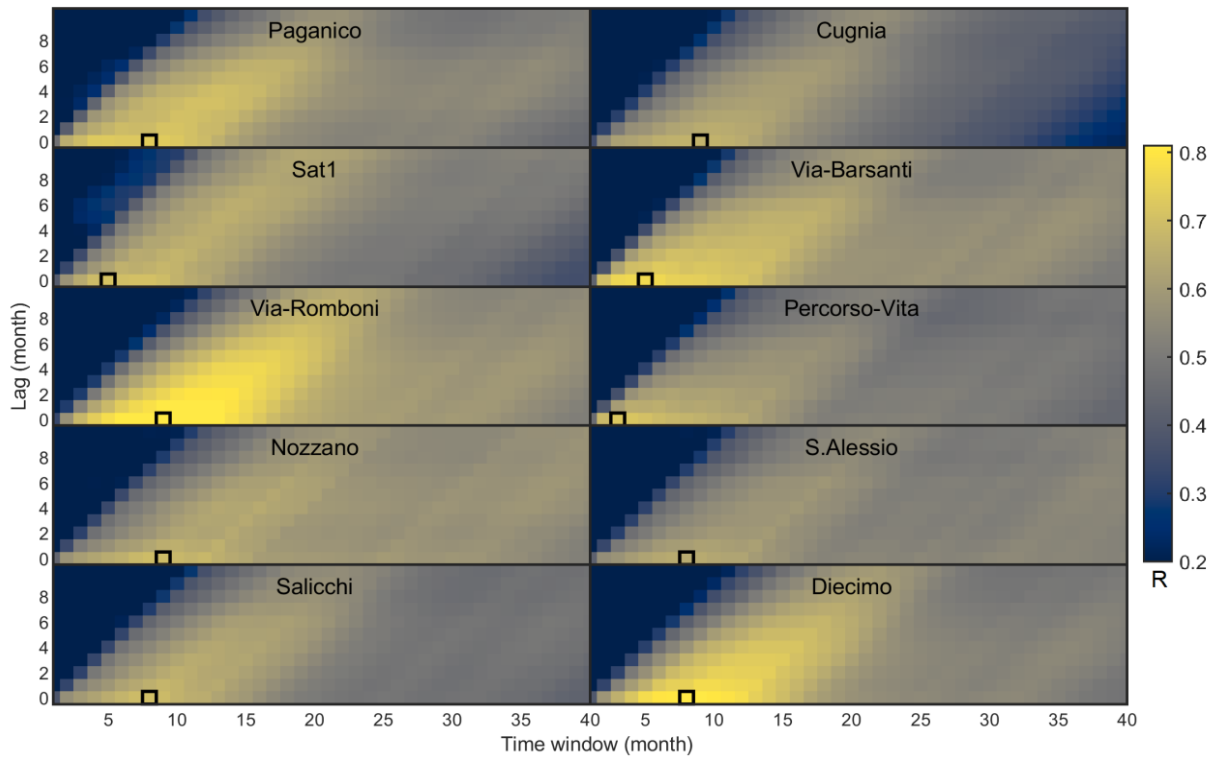

*Fig. 5 - Heat maps of the SGI-SPI correlation coefficients (R) for different time windows and lags. The black box represents the highest correlation coefficient.*

For the study area and the 10 selected wells, the precipitation accumulation periods that lead to the highest correlations do not exhibit a significant spatial variability. For all these wells but one, the maximum correlations occur considering the six- and nine-month time windows and the correlation coefficients do not considerably vary within these accumulation periods. For this reason and for clarity, in the following analysis we decided to use the SPI with six-month time window (here on denoted as SPI6) for all the 10 wells.

Once established the correlation between SPIs and SGIs, we analyzed the relationships between the two indices according to a linear regression analysis (Fig. 6). For all wells, the slope of the regression line is always lower than one; this denotes that, for the study area, in the propagation process from meteorological to groundwater droughts there is an attenuation mechanism that smooths out the negative anomalies (see e.g. Van Loon, 2015). The spread around the regression line (Fig. 6) indicates, as expected, that other factors beside the precipitation (e.g. lateral inflow/outflow, human activities) are behind the drought propagation process (Wang et al., 2016); however, the correlation between SPIs and SGIs is high and this allows us to consider this simple relationship for the subsequent analyses.

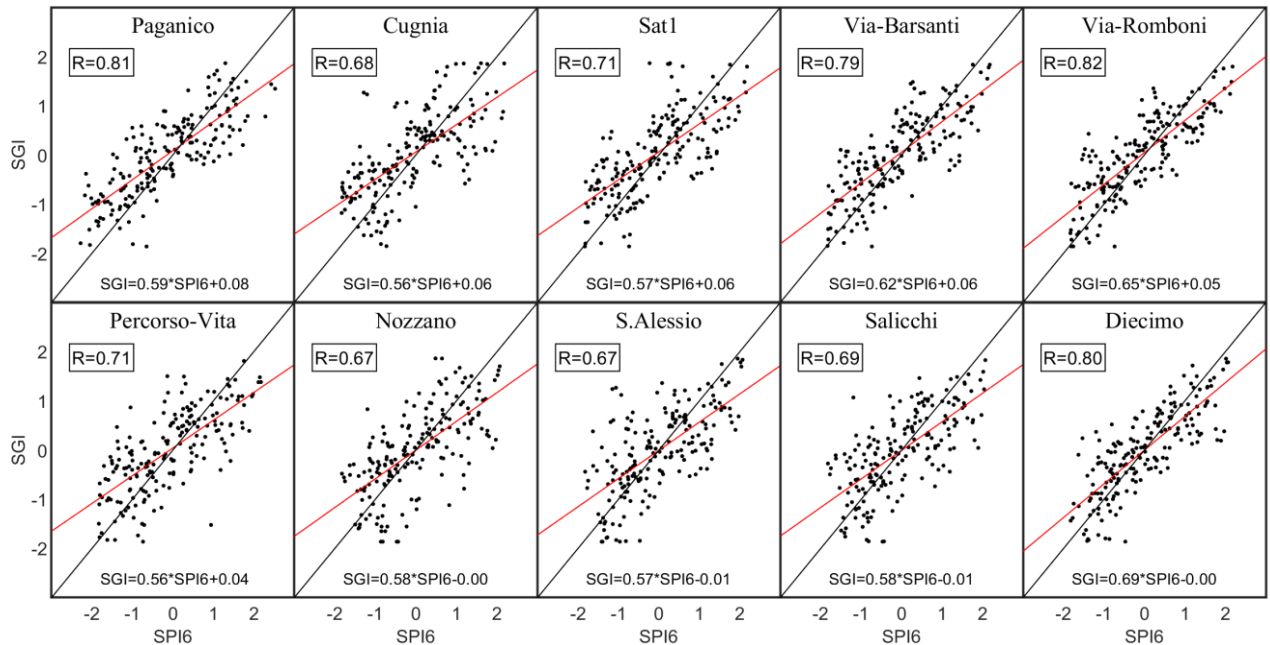

*Fig. 6 - SGIs versus SPI6; the points represent the data, the red line indicates the regression line and the black line denotes the identity line. For each well, the correlation coefficient (R) and the regression equation is reported.*

The same procedure presented above was used to investigate the correlations and relationships between SPEIs and SGIs. With reference to the wells with a correlation coefficient above the threshold (0.6), also in this case the correlations are higher considering the accumulation periods of 6, 9 and 12 months (Fig. 7). The same 10 wells, identified using SPI, satisfy the threshold condition. In general, the correlations between SPEIs and SGIs result moderately lower than those obtained processing SPIs and SGIs. In the majority of cases, the 9-month time window provides the better results, with correlation coefficients similar to those of the two adjacent accumulation periods. For this reason and for clarity, the further analyses were carried out with reference to the SPEI with a 9-month time window (here referred to as SPEI9), weighted on the four basins. An investigation on the influence of time delays between SPEIs and SGIs showed that the maximum correlations are achieved again with zero-lag for all the 10 wells (the figure is not shown for brevity). For all wells, the slopes of the regression lines are lower than the corresponding ones evaluated using SPIs, therefore a greater attenuation in the drought propagation processes was found for the study area when considering also the evapotranspiration processes. Also in this case, the spread around the regression line (Fig. 8) highlights that other factors besides the useful precipitation influence the groundwater levels; however, the high correlation between SPEIs and SGIs allows using this simple relationship for the subsequent analyses.

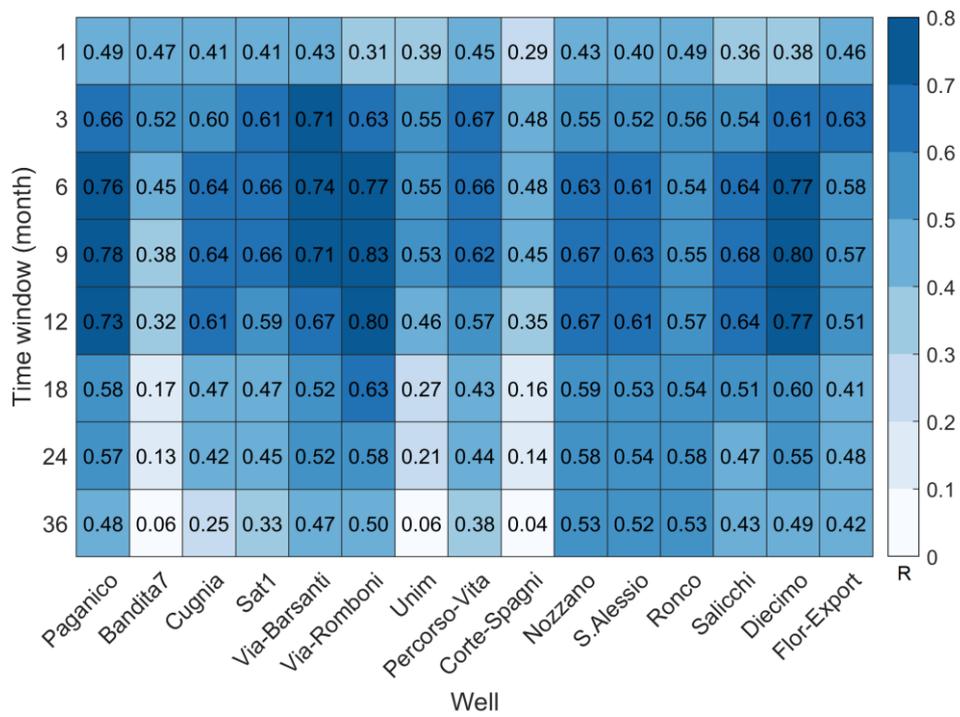

*Fig. 7 - SGI-SPEI Pearson correlation coefficients.*

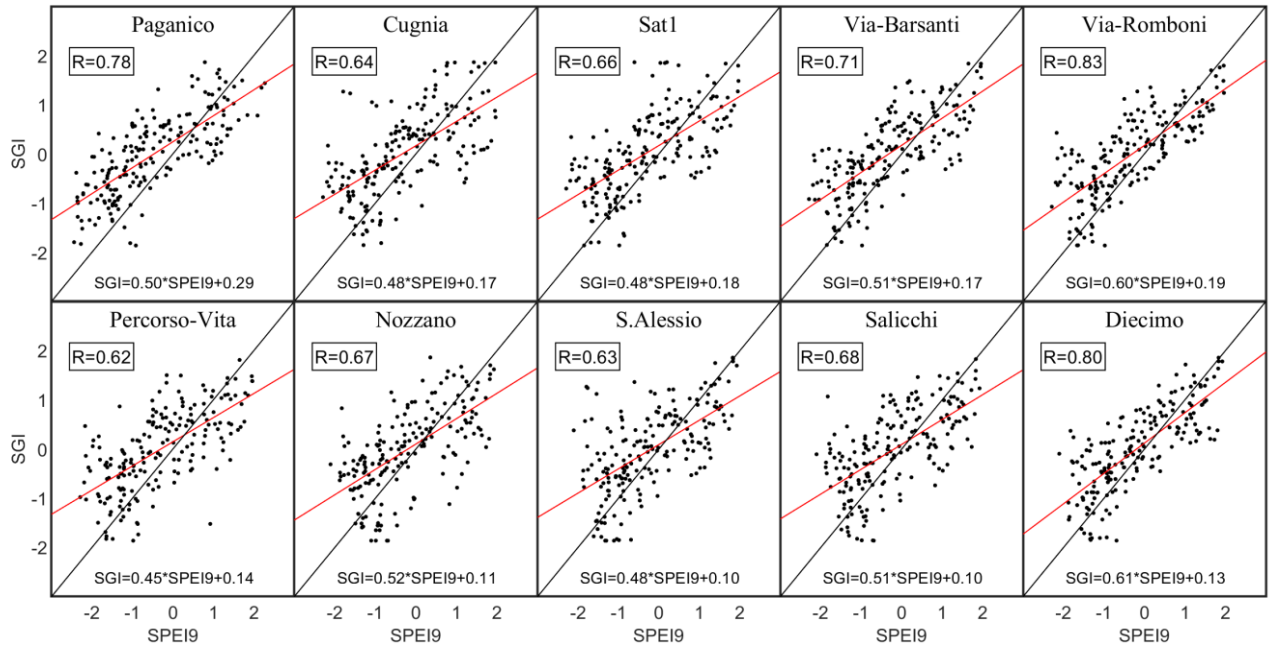

*Fig. 8 - SGIs versus SPEI9; the points represent the data, the red line indicates the regression line and the black line denotes the identity line. For each well, the correlation coefficient (R) and the regression equation is reported.*

## 3.3 Climate projections and future meteorological indices

We used an ensemble of GCM-RCM projections, downscaled and bias corrected at each station location, to represent the future climate over the study area. Even if local heterogeneities are expected in the future projections, for the sake of brevity and to have an overview of the forecasted changes in climate, we report in Fig. 9 the annual precipitation and the annual mean temperature weighted over the entire study area, for both the historical and the future periods. The data are presented in term of 10-year moving average to smooth out the natural variability and highlight the climate change components. According to both the RCP 4.5 and RCP 8.5 scenarios and the median values, the annual precipitation does not present appreciable modifications in the future for both scenarios (Fig. 9a). The variability between models is high, pointing out a large uncertainty in the future estimation of the precipitation. As for the temperature (Fig. 9b), an evident and increasing trend is detected for the future and for both scenarios. Both the historical and climate model data show that around the '90s the temperature began to increase. A similar upward trend is expected until around 2040 for both the RCPs; after this period, RCP 8.5 indicates a greater warming of the study area. Looking at Fig. 9, it can be expected that in the future, even if the precipitation does not exhibit remarkable trends, the recharge of the aquifers could be reduced due to increasing evapotranspiration phenomena triggered by the temperature rise. This endorses the importance of using meteorological indices that take into account both precipitation and temperature variables, such as SPEI, for assessing the impact of climate change on groundwater resources.

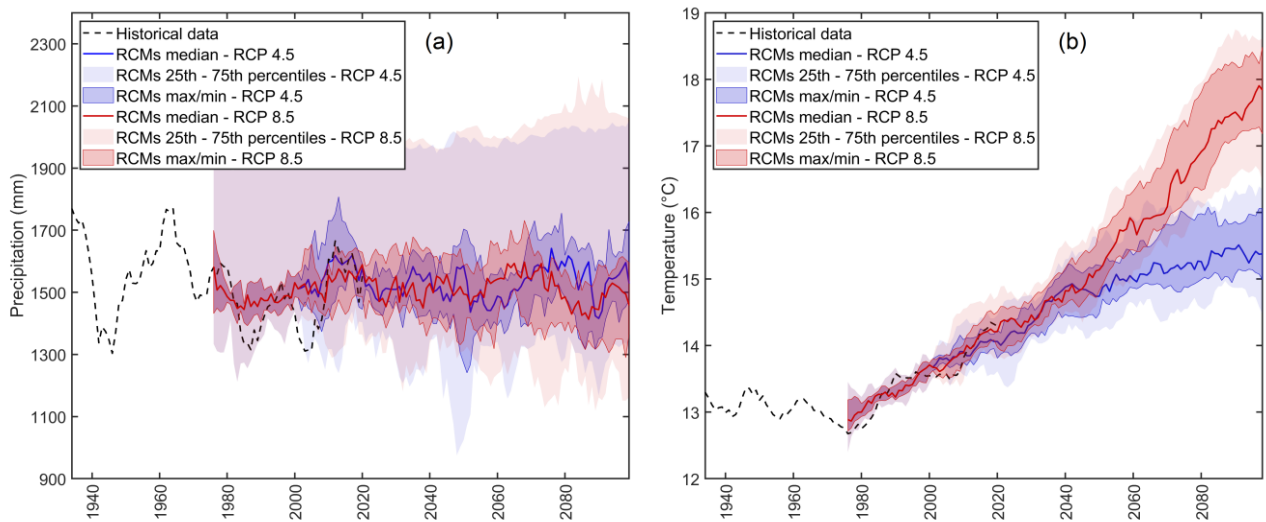

*Fig. 9 - Total annual precipitation (a) and annual average of the mean daily temperature (b) in terms of 10-year moving average observed and forecasted by the RCMs under the RCP 4.5 and RCP 8.5 scenarios. Average values over the entire study area.*

The climate model data were then used to obtain the meteorological indices as reported in Subsection 2.2.4. Before using the meteorological indices calculated from the climate models for future analysis, it is important to evaluate the reliability of the RCMs in reproducing the historical SPIs and SPEIs. We made use of the two sample Kolmogorov-Smirnov test to compare the historical and RCM meteorological indices. Since the distribution mapping procedure has been applied as bias correction method (Teutschbein and Seibert, 2012), the congruence is guaranteed at the single month scale, but for longer time windows, this may not be assured. With reference to the SPIs and a significance level of 5%, almost all samples passed the test with very few exceptions (1%) that resulted in a p-value slightly below the threshold one. For the SPEIs the percentage increases (20%) but still remains low. The results of the Kolmogorov-Smirnov test confirm that SPIs and SPEIs evaluated by the climate model data can be considered reliable.

## 3.4 Future SGIs

The SPI6 and SPEI9 values obtained from the climate models at each station location were averaged over each basin, for each RCP scenario. Making use of these values, the relationship showed in Fig. 6 and Fig. 8 were then applied to estimate the SGIs in the historical and future periods. In order to estimate the SGIs, the time series provided by the 13 RCMs were put together to constitute a single data set. In this way, the 13 realizations of the climate models have been considered equally reliable assuming that they are statistical realizations of the same stochastic process. Subsequently, we will refer to this dataset as "whole RCM ensemble".

For all wells, considering the SGI-SPI6 relationships, the CDFs of the SGIs obtained by the whole RCM ensemble, denote slight modifications with respect to the historical dataset, for both the RCP 4.5 and RCP 8.5 scenarios. Only at medium-term for the RCP 4.5 and at long-term for the RCP 8.5, a slight increase of the frequency of low SGI values has been detected. On the other hand, applying the SGI-SPEI9 relationships the CDFs of the SGIs for the future periods remarkably change with respect to the historical period: for both RCP scenarios the reduction of the median SGI values is especially pronounced at medium- and long-term.

As an example, Fig. 10 shows the empirical cumulative distribution functions of the SGIs in the historical and future periods obtained for the Paganico well (Arno portion basin) under the RCP 8.5 scenario. The envelope curves of the different CDFs obtained by considering each climate model separately show a marked uncertainty due to the differences in the individual models; this aspect is more evident in the long-term. The results for the Paganico well are summarized in Fig. 11 by means of box-whisker plots. Applying the SGI-SPI6 regression relationships, no remarkable modifications can be detected between the historical period and the future ones, the median value remains close to zero in all periods. It is noteworthy to point out that there are positive outliers due to the results of a specific model which, unlike the other RCMs, forecast abundant precipitation in the future periods. According to the SGI-SPEI9 regression relationships, a systematic reduction of the SGIs, especially at a medium- and long-term can be detected. Considering the temperature, the effects of the model with abundant precipitation are mitigated, and on the contrary there is an increase of the negative outliers.

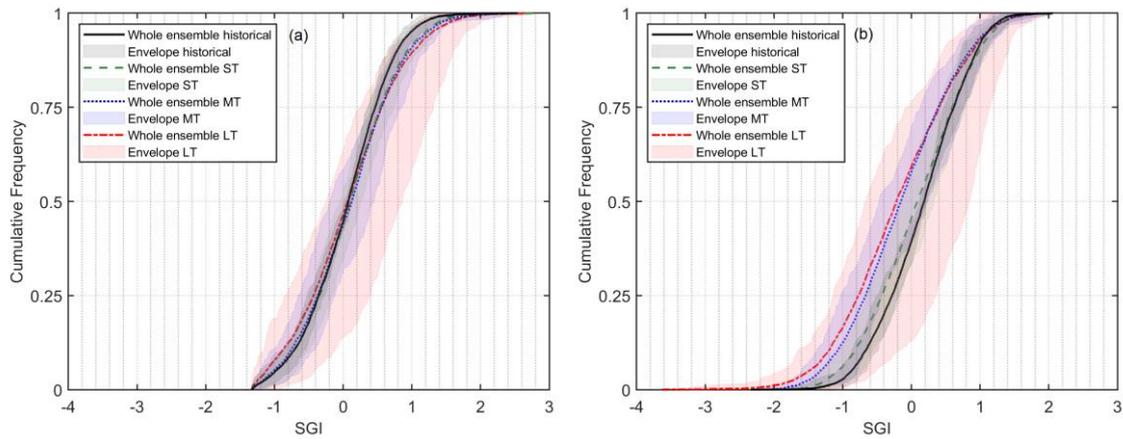

Fig. 10 - Cumulative probability distributions according to the whole RCM ensemble obtained for the Paganico monitoring well through the SGI-SPI6 (a) and the SGI-SPEI9 (b) regression equations for the historical period and at short- (ST), medium- (MT) and long-term (LT) under the RCP 8.5 scenario. Envelope curves obtained by the 13 RCM models.

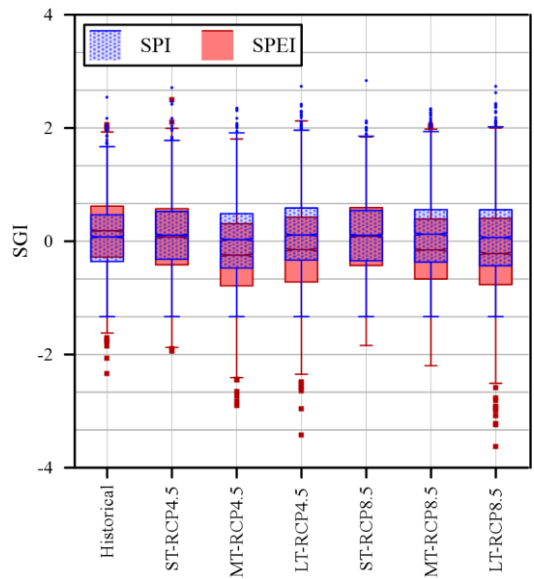

Fig. 11 - Box-plots of the SGIs obtained for the Paganico monitoring well, according to the whole RCM, through the SGI-SPI6 and SGI-SPEI9 regression equations for the historical period and at short- (ST), medium- (MT) and long-term (LT) under the two RCP scenarios. The boxplot draws points as outliers if they are greater than the mean $\pm 2.7\sigma$, where $\sigma$ is the standard deviation.

To quantify the results for all wells, some characteristic values of the SGIs defined through the SGI-SPI6 and the SGI-SPEI9 regression relationships are reported (Fig. 12). For the SGI-SPI6 relationships, looking at the 25th percentile and the median value, there is a slight decrease of the SGI in the medium-term for the RCP 4.5 and in the long-term for the RCP 8.5. Conversely, using the SGI-SPEI9 relationships, the future SGIs remarkably decrease in almost all wells. For the RCP 4.5, the medium-term period shows the greatest declines, while for the RCP 8.5 the most critical groundwater level conditions are expected in the long-term. The detected changes maintain very similar characteristics in all wells, especially within the same basin.

|   |   | SGI-SPI6 | | | | | | SGI-SPEI9 | | | | | |
|---|---|---|---|---|---|---|---|---|---|---|---|---|---|
|   |   | Historical | Differences with the historical period | | | | | Historical | Differences with the historical period | | | | |
|   |   |   | RCP 4.5 | | | RCP 8.5 | | | RCP 4.5 | | | RCP 8.5 | | |
|   |   |   | ST | MT | LT | ST | MT | LT |   | ST | MT | LT | ST | MT | LT |
| Paganico | 25th | -0.36 | 0.04 | -0.12 | 0.03 | 0.01 | -0.01 | -0.07 | -0.28 | -0.13 | -0.51 | -0.44 | -0.15 | -0.39 | -0.49 |
| Paganico | 50th | 0.08 | 0.03 | -0.05 | 0.03 | 0.03 | 0.05 | -0.01 | 0.19 | -0.12 | -0.43 | -0.34 | -0.10 | -0.34 | -0.40 |
| Paganico | 75th | 0.47 | 0.06 | 0.02 | 0.12 | 0.07 | 0.09 | 0.09 | 0.62 | -0.05 | -0.31 | -0.20 | -0.03 | -0.23 | -0.22 |
| Cugnia | 25th | -0.33 | -0.08 | -0.21 | -0.12 | -0.09 | -0.12 | -0.17 | -0.32 | -0.08 | -0.29 | -0.25 | -0.08 | -0.21 | -0.27 |
| Cugnia | 50th | 0.01 | 0.03 | -0.04 | 0.04 | 0.01 | 0.02 | -0.01 | 0.02 | -0.04 | -0.21 | -0.16 | -0.03 | -0.17 | -0.19 |
| Cugnia | 75th | 0.33 | 0.15 | 0.13 | 0.20 | 0.15 | 0.15 | 0.17 | 0.38 | 0.01 | -0.14 | -0.07 | 0.01 | -0.13 | -0.10 |
| SAT 1 | 25th | -0.41 | 0.01 | -0.12 | -0.03 | -0.01 | -0.04 | -0.08 | -0.30 | -0.08 | -0.28 | -0.24 | -0.08 | -0.21 | -0.27 |
| SAT 1 | 50th | 0.01 | 0.04 | -0.03 | 0.05 | 0.02 | 0.02 | -0.01 | 0.02 | -0.04 | -0.21 | -0.16 | -0.03 | -0.17 | -0.19 |
| SAT 1 | 75th | 0.41 | 0.07 | 0.06 | 0.13 | 0.08 | 0.08 | 0.10 | 0.37 | 0.00 | -0.14 | -0.07 | 0.00 | -0.13 | -0.10 |
| Via Barsanti | 25th | -0.46 | 0.01 | -0.13 | -0.04 | -0.01 | -0.04 | -0.09 | -0.10 | -0.08 | -0.29 | -0.25 | -0.08 | -0.22 | -0.28 |
| Via Barsanti | 50th | 0.00 | 0.04 | -0.03 | 0.05 | 0.02 | 0.02 | -0.01 | 0.25 | -0.04 | -0.22 | -0.17 | -0.03 | -0.17 | -0.20 |
| Via Barsanti | 75th | 0.44 | 0.08 | 0.06 | 0.14 | 0.09 | 0.09 | 0.11 | 0.62 | 0.00 | -0.14 | -0.07 | 0.00 | -0.13 | -0.11 |
| Via Romboni | 25th | -0.49 | -0.15 | -0.05 | -0.01 | -0.06 | -0.11 | -0.06 | -0.13 | -0.10 | -0.35 | -0.29 | -0.10 | -0.25 | -0.33 |
| Via Romboni | 50th | 0.03 | -0.08 | 0.01 | 0.02 | -0.02 | -0.05 | -0.04 | 0.28 | -0.05 | -0.26 | -0.20 | -0.04 | -0.21 | -0.23 |
| Via Romboni | 75th | 0.53 | -0.02 | 0.07 | 0.04 | 0.01 | 0.03 | 0.01 | 0.71 | 0.01 | -0.17 | -0.09 | 0.01 | -0.16 | -0.13 |
| Percorso Vita | 25th | -0.43 | 0.01 | -0.12 | -0.03 | -0.01 | -0.04 | -0.08 | -0.17 | -0.07 | -0.27 | -0.23 | -0.07 | -0.20 | -0.25 |
| Percorso Vita | 50th | -0.01 | 0.04 | -0.03 | 0.05 | 0.02 | 0.02 | -0.01 | 0.14 | -0.04 | -0.20 | -0.15 | -0.03 | -0.16 | -0.18 |
| Percorso Vita | 75th | 0.39 | 0.07 | 0.06 | 0.13 | 0.08 | 0.08 | 0.10 | 0.47 | 0.00 | -0.13 | -0.07 | 0.00 | -0.12 | -0.10 |
| Nozzano | 25th | -0.40 | 0.01 | -0.14 | -0.04 | -0.01 | -0.04 | -0.12 | -0.18 | -0.09 | -0.37 | -0.30 | -0.11 | -0.27 | -0.36 |
| Nozzano | 50th | -0.01 | 0.02 | -0.05 | 0.01 | 0.01 | -0.01 | -0.05 | 0.18 | -0.06 | -0.26 | -0.20 | -0.05 | -0.22 | -0.26 |
| Nozzano | 75th | 0.35 | 0.04 | 0.02 | 0.07 | 0.06 | 0.04 | 0.03 | 0.53 | -0.03 | -0.17 | -0.10 | 0.00 | -0.15 | -0.15 |
| S. Alessio | 25th | -0.38 | 0.01 | -0.14 | -0.04 | -0.02 | -0.04 | -0.12 | -0.17 | -0.08 | -0.33 | -0.27 | -0.10 | -0.24 | -0.32 |
| S. Alessio | 50th | 0.03 | -0.01 | -0.08 | -0.02 | -0.03 | -0.04 | -0.08 | 0.15 | -0.06 | -0.23 | -0.18 | -0.04 | -0.19 | -0.23 |
| S. Alessio | 75th | 0.40 | -0.02 | -0.03 | 0.02 | 0.00 | -0.01 | -0.02 | 0.47 | -0.02 | -0.15 | -0.09 | 0.00 | -0.13 | -0.13 |
| Salicchi | 25th | -0.39 | 0.01 | -0.14 | -0.04 | -0.01 | -0.04 | -0.12 | -0.06 | -0.21 | -0.47 | -0.41 | -0.23 | -0.39 | -0.47 |
| Salicchi | 50th | 0.00 | 0.02 | -0.05 | 0.01 | 0.00 | -0.01 | -0.05 | 0.28 | -0.17 | -0.36 | -0.31 | -0.15 | -0.32 | -0.36 |
| Salicchi | 75th | 0.36 | 0.04 | 0.02 | 0.07 | 0.06 | 0.04 | 0.03 | 0.60 | -0.12 | -0.26 | -0.20 | -0.10 | -0.24 | -0.24 |
| Diecimo | 25th | -0.47 | 0.01 | -0.16 | -0.04 | -0.02 | -0.05 | -0.14 | -0.21 | -0.11 | -0.42 | -0.35 | -0.13 | -0.32 | -0.42 |
| Diecimo | 50th | -0.01 | 0.03 | -0.06 | 0.01 | 0.01 | -0.01 | -0.06 | 0.20 | -0.07 | -0.30 | -0.23 | -0.05 | -0.25 | -0.30 |
| Diecimo | 75th | 0.42 | 0.04 | 0.03 | 0.09 | 0.07 | 0.05 | 0.04 | 0.61 | -0.03 | -0.19 | -0.12 | 0.00 | -0.17 | -0.17 |

*Fig. 12 - Differences of the median, 25th and 75th percentiles of the future SGIs with those evaluated in the historical period. The SGIs are defined through the SGI-SPI6 (left) and the SGI-SPEI9 (right) regression relationships for the historical period and at short- (ST), medium- (MT) and long-term (LT), under the RCP 4.5 and RCP 8.5.*

## 4 Discussion

A first aspect worthy of discussion concerns the relationships that represent the SGI-SPI and SGI-SPEI dependence. For the majority of wells (10 out of 15) in the study area and specific accumulation periods (6, 9 and 12 months), our results showed that the correlation coefficients are high, indicating a clear influence of the antecedent precipitations, or of the useful antecedent precipitations, on the groundwater indices. On this aspect, there is accordance with other recent studies (see e.g. Bloomfield and Marchant, 2013; Li and Rodell, 2015, Kumar et al. 2016; Van Loon et al., 2017; Uddameri et al., 2019; Guo et al., 2021).

As pointed out by Kumar et al. (2016), the propagation of a meteorological drought to the groundwater is influenced by many factors, which are related not only to the climatic characteristics but also to the basin peculiarities (such as soil properties, geology, etc.). This results in a dispersion of the observed points around the regression lines between meteorological and groundwater indices (see Fig. 6 and Fig. 8). An element to consider is that the monthly precipitation, used to evaluate SPIs and SPEIs, does not take into account in any way the intensity of the rainstorms. It is known that the water that feeds the aquifers develops with dynamics that are related to the initial soil moisture conditions and to the way in which they change during a rain event (Chow et al., 1988). If the precipitation intensity is very high, a significant portion of the volume becomes runoff and little recharges the aquifer; in the case of precipitation of modest intensity, the presence and typology of vegetation plays a fundamental role in quantifying the aquifer recharge. Even the dryness of the soil can negatively affect the infiltration rate and therefore the recharge. In addition, anthropogenic factors, such as the withdrawals for drinking or irrigation purposes, have a great relevance; moreover, they can have characteristics of marked seasonality (e.g. due to tourist presences or irrigation) that can affect groundwater levels in different ways along the year. Another source of uncertainty could be related to the presence of lateral inflow or outflow to the considered aquifers, which may affect the groundwater levels. Even with some approximations and uncertainties, all these effects can be quantified through a complete numerical modelling, which, as known, is not quick, easy and cheap to implement.

Another important issue to be considered is the accumulation period selected to compute the meteorological indices. The time window that gives the highest correlation with the SGIs can be different in relation to the examined aquifer. Several authors (Bloomfield and Marchant, 2013; Kumar et al., 2016; Soleimani Motlagh et al., 2017; Van Loon et al., 2017; Todaro et al., 2018) believe that these variations are due to the different characteristics of the aquifers under considerations: for example, the type of natural recharge (precipitation or recharge from contiguous aquifer or from a lake or stream) and its conditions (i.e. distance between the ground level and the water table). Also in this study, the SPI and SPEI time windows that provide the optimal correlations with the SGIs are not always the same for all wells, but the variation of the correlation coefficients, for accumulation periods between three and 12 months, is small. This, on the one hand, makes the selection of the optimal accumulation window more difficult; on the other hand, it justifies the choice of a single aggregation period for the entire study area. This behavior is mainly related to the characteristics of the analyzed groundwater systems; in all cases they are aquifers with phreatic surface at modest depth below to the ground surface.

An element of originality of this work is the application of an easy and fast method to assess the possible effects of climate change on the quantitative status of groundwater, combining the historical relationships between meteorological and groundwater indices with future climate projections. To achieve this result, the regression relationships between SGIs and SPIs and SGIs and SPEIs need to be considered valid also for the future. There is some debate about the reliability of using these regression relationships for future predictions. The evapotranspiration mechanisms may change as the concentration of $CO_2$ in the atmosphere increases. According to Vicente-Serrano et al. (2020), the increase in atmospheric evaporative demand resulting from an increase in the radiative component and in the temperature may not necessarily lead to an intensification of the droughts. The effect can be different if the region has a humid or dry climate and can have different impacts on meteorological, hydrological and agricultural droughts. Finally, they agree that even if plants may reduce water consumption because they optimize functions due to a favorable effect of the higher concentration of carbon dioxide, the increase in temperature causes greater evaporation from water surfaces and soil. According to Bloomfield et al. (2019), evidence of this behavior can be found from some sites in the UK that present an unusually long series of observations. According to the authors, the more frequent occurrence of groundwater drought, not accompanied by a lack of precipitation and an increase in withdrawals, is due to an increase in temperature, which induces greater evaporation from the soil above the phreatic line and especially from the capillary fringe. These results lead Bloomfield et al. (2019) to state that a change in the occurring of groundwater droughts, generated by anthropogenic warming, is already detectable. Another indirect effect of the increasing temperature is the alteration of the root system. The adaptation of plants to a warming climate is discussed by different authors (Lubczynski, 2009; David et al., 2016; Eliades et al., 2018), who highlight that trees in Mediterranean regions manage to survive droughts by extending and deepening the root systems; this behavior can lead to increasing withdrawals from the aquifer or the capillary fringe. Other authors (Teuling et al., 2013; Vicente-Serrano et al., 2014; Diffenbaugh et al., 2015; Dierauer and Zhu, 2020) emphasize the need to consider the temperature in evaluating droughts indices as it leads to a significant increase in the drought severity. Therefore, the assessment of the effects of climate change that considers only the variations in precipitation is intrinsically unreliable. For this reason, it is necessary to take into account the thermal effects in detecting climate and hydrological future trends. Some authors (Bloomfield et al., 2019; Vicente-Serrano et al., 2020) highlight that in several regions no variations in the future precipitation are forecasted but modifications, essentially increments, of the temperature could be remarkable. This is particularly evident for our case study, as showed in Fig. 9. In this regard, although in our work the SPIs and SPEIs give similar results for the historical period, this behavior may not be valid for the future. As other authors pointed out (see e.g. Kumar et al., 2016), we believe that the relationships between SGIs and SPEIs are more suitable for drought studies involving global warming conditions than the SGI-SPI ones.

Another element of discussion is that different climate models can provide very different results. For this reason, it is important to consider in the analysis an ensemble of models (Jackson et al., 2015; Mascaro et al., 2018; D'Oria et al., 2018a), which helps in visualizing the uncertainty of the results. In the present study, we applied a downscaling/bias correction technique aimed at adjusting the raw outputs of the climate models so that they better represent the statistical distribution of the observed precipitation and temperature data on a monthly scale. By doing so, the historical period is well reproduced, but the disparity between models remains for the future projections and represents a major contribution to the uncertainty of the results. Analyzing Fig. 10, it is evident that the envelope of the cumulative distribution functions (CDF) of the SGIs obtained with the climate models in the future periods is widespread. In this study, one model particularly contributes to the uncertainty of the results, providing projections of abundant precipitation and, consequently, higher SGIs than the other models. However, the estimations provided by the whole RCM ensemble are in good agreement with the median and mean CDFs, justifying the choice made in the present study to consider the model projections all together as a set of realizations of the same stochastic process.

Finally, it could be interesting to verify whether different formulas to calculate the potential evapotranspiration may affect the SPEI evaluation. Concerning this, the SGI-SPEI relationships could be different from the ones obtained in this study using the Thornthwaite equation. A possible alternative is to resort directly to the climate variables (i.e. temperature and precipitation) instead of the meteorological indices. To this end, possible future works may concern the application of machine-learning algorithms to better represent the mutual dependences among groundwater levels, precipitation and temperature.

## 5   Conclusions

In this paper, we investigated the impact of climate change on groundwater drought in northern Tuscany (Italy) making use of historical and climate model data and standardized indices. To summarize, a reduction in groundwater availability should be considered for the future in the study area. In particular, the results highlighted the importance of considering temperature to assess the impact of climate change on groundwater resources and for this reason, the regression models obtained by the SGI-SPEI relationships are more suitable for the estimation of future water levels. The procedure adopted in this study can be easily extended to different areas of interest, requiring simple observed data only in terms of groundwater levels, precipitation and temperature. We recognize the inherent degree of uncertainty that we introduce adopting the historical relationships between meteorological and groundwater indices for future analyses, but this approach can be useful for a quick estimate of the quantitative status of the aquifers under climate change scenarios. This is crucial in the process of planning integrated mitigation and adaptation strategies.

### Acknowledgements

This work was developed under the scope of the InTheMED project whose funds cover the post-doc fellowship of V.T.. InTheMED is part of the PRIMA programme supported by the European Union's HORIZON 2020 research and innovation programme under grant agreement No 1923. The authors are grateful to GAIA S.p.A. for the help during the data collection phase. The authors are thankful to the anonymous reviewers for their constructive suggestions and comments, which were very helpful to improve this manuscript.